\documentclass[prl,twocolumn,amsfonts]{revtex4}
\usepackage{amsfonts}
\usepackage{amsmath}
\usepackage{bm}
\usepackage{epsf}

\newcommand{\beq}{\begin{equation}}
\newcommand{\eeq}{\end{equation}}
\newcommand{\beqa}{\begin{eqnarray}}
\newcommand{\eeqa}{\end{eqnarray}}

\newcommand{\ket}[1]{| #1    \rangle }

\newcommand{\rref}[1]{~(\ref{#1})}

\renewcommand{\mod}{\, \mathrm{mod}\,}

\begin{document}


\newcommand{\diracsl}[1]{\not\hspace{-3.0pt}#1}

\newcommand{\pin}{\psi_{\!{}_1}}
\newcommand{\psf}{\psi_{\!{}_2}}
\newcommand{\psv}[1]{\psi_{\!{}_#1}}
\newcommand{\lab}[1]{{}^{(#1)}}
\newcommand{\psub}[1]{{\cal P}_{{}_{\! \! #1}}}
\newcommand{\sst}[1]{{\scriptstyle #1}}
\newcommand{\ssst}[1]{{\scriptscriptstyle #1}}
\newcommand{\aft}{{{\scriptscriptstyle\succ}}}
\newcommand{\bef}{{{\scriptscriptstyle\prec}}}


\title{Scaling and universality of multipartite entanglement at criticality}

\author{  Alonso Botero }
\email{abotero@uniandes.edu.co} \affiliation{
    Departamento de F\'{\i}sica,
    Universidad de Los Andes,
    Apartado A\'ereo 4976,
    Bogot\'a, Colombia}
    \affiliation{ Department of
Physics and Astronomy, University of South Carolina, Columbia, SC
29208}
\author{ Benni Reznik }
\email{reznik@post.tau.ac.il}
\affiliation{ Department of Physics and Astronomy, Tel Aviv University, Tel
Aviv 69978, Israel.
       }

\date{\today}

\begin{abstract}
Using the geometric entanglement measure, we study  the scaling of
multipartite entanglement in several 1D models at criticality,
specifically the linear harmonic chain and the XY spin chain
encompassing both the Ising and XX critical models. Our results
provide convincing evidence that 1D models at criticality exhibit
a universal logarithmic scaling behavior $ \sim {c\over 12}
\log_2\! \ell $ in the multipartite entanglement per region for a
partition of the system into regions of size $\ell$, where $c$ is
the central charge of the corresponding universality class in
conformal field theory.

\end{abstract}

\maketitle

The study of the connection between quantum entanglement and the
properties of spatially extended many-body systems   such as spin
chains~\cite{Woott01,Osborne2002,osterloh02} and harmonic chains
\cite{Auden02, botero-2004-70}, has recently attracted
considerable attention.
 This connection is especially relevant for
systems near, or at, certain quantum phase transitions, where
well-studied features of criticality such as  scale-free behavior
and universality are in fact manifestations of the entanglement
properties of the underlying quantum state.

Universal entanglement signatures of criticality have by now been
well established in the case of pure, bipartite entanglement.
 Specifically, for 1D
critical systems, the entanglement entropy of a region of size
$\ell$ and its complement  is seen to follow the universal law $ S
\sim \frac{c}{3}\log_2 \ell$, where $c$ is the central charge
characterizing a corresponding universality class in conformal
field theory (CFT) in the continuum limit. First obtained for
continuous fields within the CFT framework~\cite{holzhey}, the
result has  by now been widely verified in discrete systems such
as spin chains~\cite{vidal2003prl}, harmonic
chains~\cite{botero-2004-70}, and
fermions~\cite{wolf2005,gioev2005}.
%

That many-body systems should also manifest properties of genuine
multipartite entanglement (MPE) has been evidenced, for example,
in spin chains~\cite{bruss2005,multi2,multi3,multi4}, and as shown
in the simulation of many-body ground states within the
matrix-product state   framework\cite{mps-rev}, MPE at criticality
may be highly non-trivial. Connections between critical behavior
and MPE in 1D spin models have been particularly elaborated by
 \cite{wei-2005-71,chen,oliveira1,oliveira2}, which find non-analytic behavior of MPE at criticality
 and other signs of universality.


In this Letter we further explore the MPE of critical 1D systems
by addressing a question in the spirit of renormalization that
naturally arises in this context:  Given the scale-free nature of
the continuum limit of the critical system, how does the MPE
between regions of the same size scale under coarse- or
fine-graining, that is, as we vary the size of the regions (Fig.
1)?  We investigate this problem in both harmonic and XY spin
chains, using as a measure of MPE the geometric measure of Wei and
Goldbart~\cite{wei-2003-68,wei-2005-71}. Our main finding is that
with respect to a partition of the system into equal regions of
size $\ell$, the geometric MPE per region, ${\cal E}$, shows the
logarithmic scaling behavior
\begin{equation} \label{density}
{\cal E} \sim {c^*\over 12}  \log_2 \ell \,
\end{equation}
at criticality. Here, $c^*$ is a constant that  most probably is
the central charge of the relevant universality class, as
evidenced from our numerical results and the correspondence with
the case bipartite case, for which a connection can be established
independently using previous results in the literature.

\begin{figure}
   \epsfxsize=2.5truein

\centerline{\epsffile{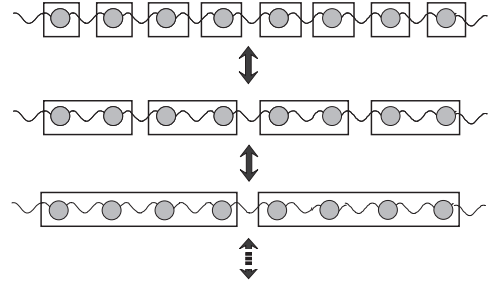}}
\medskip
\caption[ ]{Coarse graining of a system to $N$ regions of size $l$, with $Nl$ fixed. }
 \end{figure}


For an entangled state of $N$ parties, the geometric entanglement
measure~\cite{wei-2003-68} is defined as $E(\psi) = -\log
|\Lambda|^2$, where $|\Lambda|^2$ is the maximal overlap $|\langle
\psi|\phi_1 \phi_2\ldots \phi_N \rangle|^2$ over all possible
product states $|\phi_1 \phi_2\ldots \phi_N \rangle$; the optimal
states and $\Lambda$ are in turn solutions to the nonlinear
eigenvalue equation
 \begin{equation}
\label{eigeq} \langle \phi_1 \phi_2\ldots { \widehat{\phi_i }
}\ldots\phi_N |\psi\rangle = \Lambda |\phi_i\rangle \, ,
\end{equation}
where ${ \widehat{\phi_i} }$ stands for the exclusion of the state
$|\phi_i\rangle$. Here, we will be interested in the ground state
$\ket{\psi}$ of large circular chains of spins and oscillators,
with the $N$ parties representing regions of size $\ell$ each, so
that the total number of spins or oscillators is $N \ell$.
Considerable simplifications of the nonlinear eigenvalue problem
under these conditions allow for a feasible numerical
implementation of the solution: First, due to translational
invariance, the optimal state takes the form $\ket{\phi}^{\otimes
N}$, and the problem reduces to finding a single state
$\ket{\phi}$ for one region. Furthermore,
 ground states of oscillator   and spin chains in the XY
model are both describable in the language of bosonic or fermionic
gaussian states, where the connection in the XY case is through a
Jordan-Wigner transformation to effective fermion variables. Since
 partial tracing is a gaussian operation, a solution to Eq. \rref{eigeq}
 can  be found within the class of gaussian states.
As is well known, gaussian states  can be characterized by a
single covariance matrix\cite{notation} that takes the block form
\begin{equation}
\mathbf{M}_{\rm boson} = \left(%
\begin{array}{cc}
  \mathbf{G} & \mathbf{K} \\
  \mathbf{K^T} & \mathbf{H} \\
\end{array}%
\right) \, , \ \ \ \mathbf{M}_{\rm fermion} = \left(%
\begin{array}{cc}
  \mathbf{A} & \mathbf{C} \\
  -\mathbf{C}^T & \mathbf{B} \\
\end{array}%
\right)\, ,
\end{equation}
where $\mathbf{M}_{\rm boson}$ is symmetric and $\mathbf{M}_{\rm
fermion}$ is antisymmetric. Further simplification follows from
the fact  that in the cases of interest, the conditions
$\mathbf{K}=0$ and $\mathbf{A}=\mathbf{B}=0$ are satisfied. For
pure states, these conditions imply that in the bosonic case
$\mathbf{G} \mathbf{H} = \openone$, while in the fermionic case
$\mathbf{C}\mathbf{C}^T = \openone$; thus, all the information of
the gaussian state can be encoded in a single block (e.g.,
$\mathbf{G}$ or $\mathbf{C}$) . We shall therefore refer to a
single $N \ell \times N \ell$ matrix $\bm{\Omega}$ for the full
state $\ket{\psi}$ and an $\ell \times \ell$ matrix
$\bm{\omega}_{\rm opt}$ for the optimal state $\ket{\phi}$, where
both $\bm{\Omega}$ and $\bm{\omega}$ are either symmetric
(bosonic) or orthogonal (fermionic). Finally, due to translational
symmetry, $\bm{\Omega}$ has a Toeplitz form at the level of $\ell
\times \ell$ blocks, and may thus be brought through  a matrix
version of Bloch's theorem to the block-diagonal form $
\bm{\omega} = \bigoplus_{\eta} \bm{\omega}_\eta $ where the set
$\{ e^{i \eta} \}$ are the $N$th roots of unity, and the
$\bm{\omega}_\eta$ are hermitian (bosonic) or unitary (fermionic)
$\ell \times \ell$ matrices, given by
\begin{equation}
[\bm{\omega}_\eta]_{i j} = \frac{1}{N}\sum_{m =0}^{N -1}
[\bm{\Omega}]_{m { \ell} + i ,j}\, e^{ i  m \eta    } .
\end{equation}
The optimal gaussian solution to the eigenvalue problem of Eq.
\ref{eigeq} can then be recast as the equation
\begin{equation}
\label{phieq}
 \bm{\omega}_{\rm op}^{-1} = \frac{1}{N}\sum_{\eta} \frac{2}{\bm{\omega}_{\rm op}  +
\bm{\omega}_\eta } \, .
\end{equation}
With this form, a numerical computation of $\bm{\omega}_{\rm opt}$
can  be obtained iteratively starting with a trial
$\bm{\omega}_{\rm opt}$ in the right hand side and iterating until
a fixed point is reached. Our experience shows that in most cases
twenty or so iterations suffice to reach an acceptable fixed
point.

Once $\bm{\omega}_{\rm opt}$ is determined, the geometric
entanglement can  be obtained from the  sum $E(\psi) = \sum_\eta
E_\eta$, where
\begin{equation}
\label{phient}
 E_\eta(\psi) = \pm\log_2\left|\frac{\det  \frac{1}{2} \left[\bm{\omega}_{\rm op} +
 \bm{\omega}_\eta\right]
   }{  {\sqrt{\det \bm{\omega}_{\rm op}  \det \bm{\omega}_\eta } }}\right|
\, ,
\end{equation}
are the partial contributions from each sector and where the minus
sign applies to the fermionic case. The average of the partial
entanglements $E_\eta(\psi)$ gives the entanglement per region or
\emph{entanglement density} ${\cal E}(\psi) \equiv E(\psi)/N$,
which is our quantity of interest.

%
%

 We first consider the circular harmonic chain
 described by a Hamiltonian with a single parameter $0 \leq \alpha
 < 1$, $
H_\alpha = \sum {p_i^2\over 2}+{q_i^2\over 2} - \alpha q_iq_{i+1}
$, with $i$ periodic in  $N \ell$. Here, $\bm{\Omega}$ has entries
given by $[\bm{\Omega}]_{i j} = 2 g(|i-j|)$ where $g(l)$ is  the
momentum correlation function $\langle p_0 p_l \rangle$ and is
diagonalized by  discrete circular wave normal mode  functions
indexed by an angle $\theta_k = \frac{2 \pi}{N \ell} k$, with
eigenvalues given by the dispersion relation
$\omega(\theta_k)=\sqrt{1-\alpha\cos(\theta_k)}$.  For finite
$\alpha$, the correlation function decays exponentially with $l$
with   correlation length $\xi \simeq 1/\sqrt{2(1-\alpha)}$.
Criticality corresponds to the limit $\alpha\to1$, in which the
system becomes gapless, ($\omega(\theta) \propto |\theta| $  for
small theta), the correlation length diverges, and the correlation
function exhibits power-law behavior $g(l) \sim 1/l^2$.

It will be instructive to briefly develop a picture of the optimal
solution based on  the
 Bloch decomposition of $\bm \Omega$.
In the harmonic chain, the $\bm{\omega}_\eta$ are  given in terms
of the dispersion relation $\omega(\theta)$ and circular plane
waves  with shifted node numbers
\begin{equation}\label{omegaetas}
[\bm{\omega}_\eta]_{i j} = \frac{1}{ \ell}\sum_{k =0}^{\ell -1}
\omega( \theta_{k+ \nu_\eta})\,  e^{ i
{\theta}_{k+\nu_\eta}(i-j)}\,\ \ \
\end{equation}
where $\nu_\eta = (\eta \mod 2 \pi) /2 \pi$ is the winding number
of $\eta$.
We can interpret $\bm{\omega}_\eta$  as (twice) the momentum
correlation matrix $\langle \pi_i \pi_j^\dagger\rangle_\eta$ for
the vacuum state of a  complex scalar field on a linear lattice of
$\ell$ points, with Hamiltonian $H_\eta =
\frac{1}{2}\mathbf{\pi}^\dagger\cdot\mathbf{\pi} +
\frac{1}{2}\mathbf{\phi}^\dagger \mathbf{V}_\eta \mathbf{\phi}$
and potential matrix $\mathbf{V}_\eta=\bm{\omega}_\eta^2 $, where
\begin{equation}
[\mathbf{V}_\eta]_{i j} = \delta_{ij} + \frac{\alpha}{2}\left[
\delta_{|i-j|,1} +  e^{i \eta}\delta_{i,1}\delta_{j,\ell} + e^{-i
\eta}\delta_{i,\ell}\delta_{j,1}\right]\,
\end{equation}
(this result is easily derived by applying the Bloch decomposition
to  the potential matrix for the whole chain). The case $\eta=0$
corresponds to a translationally invariant closed chain of $\ell$
oscillators, while for $\eta \mod 2 \pi \neq 0$, translational
invariance is lost by the appearance of a ``twisted" coupling
between the first and last oscillators, which in the continuum
correspond to the twisted boundary conditions $\phi(0) = e^{i
\eta}\phi(\ell)$ for a complex Klein-Gordon field. Now, due to
periodicity in $\eta$, the $\bm{\omega}_\eta$ describe a closed
loop in the space of Hermitian $\ell \times \ell$ matrices, and
$\bm{\omega}_{op}$ corresponds to a generalized ``center of mass"
for this loop with respect to the distance measure~(\ref{phient}).
As $\ell$ becomes large, the perturbation to the free modes on the
circular chain becomes noticeable only at the longest wavelengths,
for which $\delta \lambda/\lambda = \nu/k $ is appreciable, and
gives rise to small $\sim 1/\ell$ corrections to the spectrum.
Thus we expect that the distances  $E_\eta$ should converge to a
common value for large $\ell$. However, it is important to note
that as $\alpha \rightarrow 1$,   the partial contribution
$E_{\eta=0}$ diverges independently of $\bm{\omega}_{\rm opt}$,
this is due to the vanishing determinant of $\bm{\omega}_{\eta
=0}$ and in turn  to the dispersion relation $\omega(\theta)=0$
for the $\theta=0$ mode at criticality. Thus, for $\eta = 0$, the
relevant distance as $\alpha \rightarrow 1$ should in fact be
taken to be $E^{(ren)}_{\eta=0} = E_{\eta=0} - E_{\rm div}$, where
$E_{\rm div}\equiv -\frac{1}{2 }\log \omega(0)$ $\simeq \frac{1}{2
}\log_2 \xi$ is the divergent contribution of the zero mode. The
renormalized distances are well-defined at criticality and are
indeed seen to converge to common values as $\ell \rightarrow
\infty$.

\begin{figure} \label{2}
   \epsfxsize=3.25truein
\centerline{\epsffile{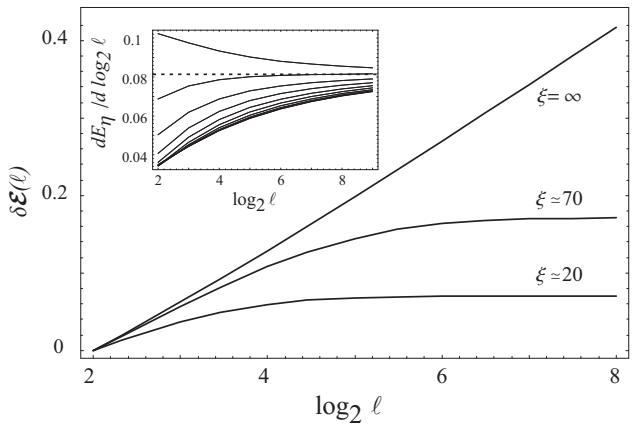}}
\medskip
\caption[ ]{Block size dependence of the geometric entanglement
per block ${\cal E}= E/N$, as measured from its value at $\ell=4$
($\delta{\cal E}(\ell)\equiv{\cal E}(\ell)-{\cal E}(4)$) vs.
$\log_2 \ell$, for three values of the correlation length $\xi$ in
the harmonic chain, for $N=10$ blocks. Inset: Convergence of the
slopes of partial contributions ${ E}_\eta$ as function of $\log_2
\ell$, for $N=20$, at criticality. Curve order (starting from
bottom curve) proportional to $|\eta-\pi|$ with $\eta \in [0,2\pi)
$; dashed line corresponds to $c/12 \simeq 0.0833$.}
\end{figure}

Turning now to our numerical results, we first note that the use
of an entanglement density is indeed appropriate, as the
(renormalized) densities  rapidly converge to $N$-independent
values at small values of $N$. This can be attributed to the fact
that increasing the density of points in the previously mentioned
closed loop of hermitian matrices, has only minor effects on the
location of the ``center of mass". We also  comment on the optimum
matrix $\bm{\omega}_{op}$:  the corresponding potential matrix
$\bm{V}_{op} = \bm{\omega}_{op}^2$ was found to have, up to small
fluctuations, the same structure of that of a chain of length
$\ell$ with the same value of $\alpha$ but with open boundary
conditions; thus, in a first approximation, the optimum potential
matrix appears to be the average of the $\bm{V}_{\eta}$ matrices
as one may expect from the center of mass
 picture. Turning then to the
entanglement density as a function of $\ell$, in Fig.~2 we present
our results  for different correlation lengths $\xi$. As can be
seen, the increase in entanglement for non-critical values
saturates when $\ell \sim \xi$. This limiting behavior can be
attributed to the entanglement of oscillators within a distance
$\sim \xi$ from the interfaces between regions, in which case the
total entanglement is  determined only by the number of
partitions. On the other hand, at criticality $\xi = \infty$, the
renormalized entanglement per region is found to scale
 with $\ell$ as
\begin{equation}\label{c'}
{\cal{E}}^{(ren)}(\psi) = \kappa^* \log_2{\ell}
\end{equation}
where $\kappa^*$ is a coefficient that one may  expect to depend
on the CFT central charge $c=1$, as does  the bipartite
entanglement entropy.  For the finite values considered in our
computations, the coefficient $\kappa^*$ in fact varies slowly as
$\ell$ is increased and it therefore becomes difficult to infer
the limiting value from the total entanglement density. However,
an examination of the instant slopes of the partial contributions
$ E_\eta $ reveals (as shown in Fig. \ref{2}) that they all
converge to a single limiting value, from both directions. We have
estimated this limiting value by fitting  the instantaneous slopes
to the function $ \kappa_\nu(\ell) = \kappa^* +
A(\nu)\ell^{-\lambda}$, where $A(\nu)$
 is quadratic in $\nu$ and symmetric about $ \nu =1/2 $ (from periodicity),  for  various values of
partitions up to $N=100$, and  obtain that $\kappa^*=0.0837$ with
a $5\% $ error, independently of the number $N$ of partitions.
Within the margin of error, this value is consistent with
$\kappa^*=c/12$ where $c$ is the central charge  $c=1$ of the
bosonic CFT universality class.

\begin{figure} \label{3}
   \epsfxsize=3.25truein
\centerline{\epsffile{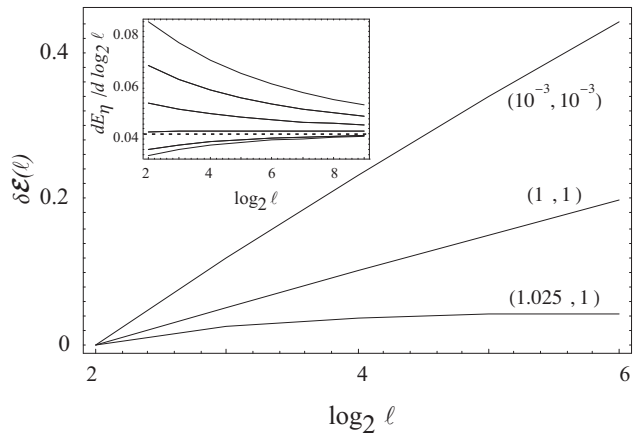}}
\medskip
\caption[ ]{  Block size dependence of the geometric entanglement
per block ${\cal E}= E/N$, as measured from its value at $\ell=4$
($\delta{\cal E}(\ell)\equiv{\cal E}(\ell)-{\cal E}(4)$) vs.
$\log_2 \ell$, for three values of $(\gamma, \lambda )$ on the
phase plane in the $XY$ model, for $N=20$ blocks. From top to
bottom: vicinity of the bosonic critical line ($XX$ model),
fermionic critical line (Ising), and a non critical value showing
saturation. Inset: Convergence of the   partial contributions ${
E}_\eta$ as function of $\log_2 \ell$, for $N=10$, at the Ising
model critical point. Curve heights follow same order as in
Fig\ref{2}; dashed line corresponds to $c/12 \simeq 0.0417$ . }
 \end{figure}


To test the universality of our results, we consider the $XY$
  spin chain model, which includes two
  critical regions associated with different CFT universality classes.
  The XY Hamiltonian takes the form $
H_{\lambda,\gamma} =- \sum \lambda Z_i +{1+\gamma \over
2}X_iX_{i+1} + {1-\gamma \over 2} Y_i Y_{i+1} $, where
$X_i,Y_i,Z_i$ are Pauli matrices at the site $i$, and the
anisotropy $\gamma$ and the field strength $\lambda$ parametrize
the ground state phase diagram. The critical lines are $\gamma=0$,
 associated with  the bosonic
($c=1$) universality class, and $\lambda=1$, associated with  the
fermionic ($c=1/2$)  class. By means of a Jordan-Wigner
transformation, $H_{\lambda,\gamma}$ can be transformed to a free
fermions system,  the ground state of which is gaussian fermionic
and can be characterized by an orthogonal matrix $\bm \Omega$ with
eigenvalues
\begin{equation}
\omega( \theta ) = \frac{ (\cos(\theta) - \lambda) - i \gamma
\sin(\theta) }{\sqrt{ (\cos(\theta) - \lambda)^2 + \gamma^2
\sin(\theta) }}
\end{equation}
and circular plane waves with momentum label $\theta$ as
eigenstates. The corresponding $\bm{\omega}_\eta$ matrices can be
obtained by using this expression for $\omega( \theta )$ in
Equation \ref{omegaetas}.

As depicted in Figure 3, the XY model also shows saturation for
non- critical regions and logarithmic scaling behavior for spin
chain models in at criticality.  For the critical Ising point,
$\gamma= \lambda =1$, very good agreement is obtained with
$\kappa^* \sim 0.043$, consistent with $c/12$ for $c=1/2$.  In the
critical $XX$ case, $\lambda=0$, $ 0\le\gamma<1$, our iteration
scheme converges slower, and large $\ell$ values become harder to
obtain numerically due to precision losses at every iteration.
Still, logarithmic scaling was evidenced with $\kappa^*$
consistent with $c/12$ for $c=1$ within  $~15\%$.

The slope value $c/12$ is consistent with the results for $N=2$,
which are analytically tractable, owing to the fact that in the
bipartite case  the geometric entanglement  is  the logarithm of
the largest Schmidt coefficient  of the state. As discussed in
~\cite{botero2003,fermionmodewise}, with respect to any bi-partite
split the ground state of a quadratic boson or fermion hamiltonian
can always be expressed as the product of two-mode entangled
states of the form $ (1\pm e^{-\beta})^{\pm 1/2}\sum_n e^{-\beta
n}|n,n\rangle$ where the $\beta's$ are related to the symplectic
eigenvalues of the reduced covariance matrix and $n$ goes from
zero to $\infty$  for bosons,  or $1$ for fermions.  The geometric
entanglement is therefore the ``free energy" $E =\pm\sum_\beta
\log(1\pm e^{-\beta})$ coming from the $n=0$ coefficients. In the
continuum limit this quantity is equal to half the von Neumann
entropy, if the density of modes is constant  for the range of
contributing modes  $\beta \lesssim 1$, as occurs at criticality
(a rigorous derivation is found in~\cite{Orus06}). Using the
result of Calabrese and Cardy \cite{calabrese-2004-0406}, where
for a symmetric split of a chain of size $2 \ell$, the
entanglement entropy is found to scale like $S \sim
\frac{c}{3}\log \ell$, the resulting geometric entanglement
density is then found to be $ {\cal E}_{N=2} = \frac{1}{2} E =
\frac{1}{4} S $, and thus to scale  as $\sim \frac{c}{12} \log
\ell$.

Elsewhere, Bravyi\cite{bravyi} has proposed a generalization of
the entanglement entropy for multipartite states, corresponding to
the minimum attainable Shannon entropy of the joint measurement
outcome distribution when considering all possible multilocal
measurement bases. Since the overlap between an entangled state
$\ket{\psi}$  and any product state cannot exceed in magnitude the
maximal value $|\Lambda|$ from eq.~\rref{eigeq}, the Shannon
entropy of the squares of the coefficients in any product basis
expansion of the state cannot exceed the entropy $-\log_2
|\Lambda|^2$ of an equally-weighted superposition of
$1/|\Lambda|^2$ product-state terms. The geometric entropy is
therefore  a lower bound on the multipartite entanglement entropy.
Roughly speaking, such an entropy may be considered a measure of
the effective Schmidt number~\cite{schmidtmeasure}, or number of
terms in a multilocal orthogonal decomposition of the state. Our
result therefore suggests that at criticality this effective
number must scale with the region size $\ell$ no slower than $\sim
\ell^{\frac{N c}{12}}$. An interesting open question is the extent
to which the exponent $\frac{N c}{12}$ is a good characterization
of the actual effective Schmidt number of 1D ground states for
critical systems.

To conclude, in the present work we found convincing evidence that
multi-partite entanglement in 1D systems, manifests at criticality a
logarithmic scaling behavior as well as universality; properties
which have so far been established only for bi-partite entanglement
at criticality. This finding, has been here demonstrated for free
Gaussian harmonic and spin chain models, and will hopefully be
further tested in other discrete solvable models, as well as in the
framework of conformal field theory. It would be also very
interesting to formulate the problem at hand in conformal field
theory, since the connection of the present result with some
properties of CFT is at present an intriguing open question. Finally
we hope that, the emerging understanding of the properties and
structure of entanglement in many-body systems, will also turn
helpful in further developing tools
for studying many-body systems.


A.B. acknowledges support from Colciencias, contract RC-2005-2003,
B.R. acknowledges the Israel Science Foundation, grant 784/06.


\begin{thebibliography}{99}
\bibitem{notation}
In terms of  $ x_i = a_i + a_i^\dagger $, $ y_i
= i (a_i^\dagger - a_i)$ for fermions, and $ x_i = (a_i +
a_i^\dagger)/\sqrt{2} $, $ y_i = (a_i-a_i^\dagger)/\sqrt{2}i$ for
bosons, and $\eta=(x_1,x_2,\cdots y_1,y_2\cdots)$ we have $[M_{\rm
fermion}]_{\alpha,\beta} =  {1\over 2i} \langle
[\eta_\alpha,\eta_\beta]\rangle$ and $[M_{\rm
boson}]_{\alpha,\beta} = {1\over 2} \langle
\lbrace\eta_\alpha,\eta_\beta\rbrace\rangle$.
\end{thebibliography}

\end{document}